# A Bibliometric Model for Identifying Emerging Research Topics[*]

Qi Wang

[1]Danish Centre for Studies in Research & Research Policy, Aarhus University, Denmark

[2]INDEK, KTH-Royal Institute of Technology, Sweden

Email：qiwang@ps.au.dk

**Abstract**

Detecting emerging research topics is essential, not only for research agencies but also for individual researchers. Previous studies have created various bibliographic indicators for the identification of emerging research topics. However, as indicated by Rotolo et al. (2015), the most serious problems are the lack of an acknowledged definition of emergence and incomplete elaboration of the linkages between the definitions that are used and the indicators that are created. With these issues in mind, this study first adjusts the definition of an emerging technology that Rotolo et al. (2015) have proposed in order to accommodate the analysis. Next, a set of criteria for the identification of emerging topics is proposed according to the adjusted definition and attributes of emergence. By the use of two sets of parameter values, several emerging research topics are identified. Finally, evaluation tests are conducted by demonstration of the proposed approach and comparison with previous studies. The strength of the present methodology lies in the fact that it is fully transparent, straightforward, and flexible.

**Keywords:** emerging research topics, identification, bibliometric analysis, definition, attributions

**Introduction**

Detecting emerging research topics is useful for research foundations and policy makers aiming to promote and enhance the development of potentially promising research topics. Evidence for the importance of identifying emerging research topics can be found in recent research projects. For instance, the European Research Council (ERC) supported Emerging Research Areas and their Coverage by ERC-Supported Projects (ERACEP) in order to identify emerging areas by topic and to analyze the extent to which ERC-funded projects contributed to these emerging areas in 2009 (Reiss et al., 2013). In addition, as discussed by Small et al. (2014), the Intelligence Advanced Research Projects Activity (IARPA) funded Foresight and Understanding from Scientific Exposition (FUSE) in 2011, one of the main purposes of which is to nominate emerging technologies.

In the light of current policy orientations, numerous studies have proposed definitions and methodologies designed to comprehend, define, and detect emerging topics (e.g., Glänzel & Thijs, 2012; Small, 2006; Small et al., 2014). However, Cozzens et al. (2010) point out that most studies

---

actually focus on the measurement of emerging research topics rather than the identification of these topics. In terms of measurement, they observe that "most of the quantitative studies of emerging technologies have taken the results of qualitative processes as a starting point, and characterized pre-identified areas quantitatively" (p. 367). While these types of studies are useful for understanding the attributes of an emerging topic, they cannot satisfy the requirements of policy makers and funding agencies, who are more interested in discovering and revealing emerging research topics. However, only a few studies have proposed methodologies that aim to fulfill this purpose (e.g., Reiss et al., 2013; Small et al., 2014).

Regarding the definition of emergence, a recent paper by Rotolo et al. (2015) systematically reviews various relevant definitions and major empirical approaches for the measurement and identification of emerging technologies. These researchers acknowledge that although many concepts exist, the fundamental attributes of an emerging technology remain ambiguous, and the connections between definitions and the approaches that have been created are fragile. They further indicate that while a wide variety of bibliometric indicators have been developed, bibliometric methods for the identification of emerging technologies generally "lack strong connections to well thought out concepts that one is attempting to measure" (p.1827). This conclusion is in line with those of Cozzens et al. (2010) and Small et al. (2014), who also indicate that the concept of emergence is seldom defined even though the term is widely used. The present study is in agreement with the opinions that the definition and attributes of an emerging topic are still ambiguous and that linkages between concepts and proposed indicators for operationalization are not well established. This lack of clarity is in fact a general limitation of studies that intend to identify emerging topics.

In order to address this limitation, this study first adjusts the definition of an emerging technology that Rotolo et al. (2015) have proposed and develops a nuanced conception of what constitutes an emerging research topic. Next, in accordance with this definition and carefully delineated attributes of emergence, a set of criteria for the identification of emerging topics is proposed. This new approach is applied to research topics that are established using direct citation relations of individual publications by employing two sets of parameter values, and hence, several emerging topics are identified. Finally, evaluation tests are conducted by demonstrating the proposed approach and comparing it with the results from previous studies. The present study is more focused on the construction of connections between the concept of an emerging research topic and the indicators that have been created for identifying these topics.

**Related Work**

Previous studies have proposed many methods for identifying emerging research topics, including both qualitative and quantitative measures. Some have focused on the identification of emerging technologies (e.g., Cozzens et al., 2010; Rotolo et al., 2015) rather than of emerging research topics, as this study does on topics relating to science. While there is some overlap between technologies and research topics, they remain distinct concepts. Rotolo et al. (2015) point out that although the importance of science for the development of technologies is widely acknowledged, "not all technological revolutions may depend on breakthrough advances in science" (p. 1832). The discussion presented here is not, however, confined to emerging research topics; some studies on the identification of emerging technologies are also reviewed. Further, our focus is on the studies that have used quantitative approaches for the detection of emergence, especially those that are bibliometric-based. Within this scope, previous studies generally consist of two steps: constructing clusters of individual publications on the basis of the direct citations among them, and then creating criteria to identify emerging research topics.

*Establishing Research Topics*

First, it should be mentioned that journal classification systems, such as Web of Science (WoS) and Scopus, are always coarse-grained. It is difficult to observe the dynamics of science using such systems. For this reason, most studies on the identification of emerging topics establish a fine-grained classification system at the outset.

Various types of relations among publications can be used to establish research topics, such as citation-based, text-based, and hybrid approaches. In the case of text-based approaches, some studies have established clusters based on the co-occurrence of terms (e.g., Furukawa et al., 2015; Lee, 2008; see also Rotolo et al., 2015, pp. 1834). In the case of citation-based approaches, different citation relations, for instance direct citation relations (e.g., Kajikawa & Takeda, 2008; Kajikawa et al., 2008; Shibata et al, 2008, 2011), bibliographic coupling relations (e.g. Morris et al., 2003), and co-citation relations (e.g. Small, 2006; Upham & Small, 2010), can all be used to aggregate publications. Furthermore, Small et al. (2014) have combined co-citation and direct citation clustering methods into a large-scale dataset for detecting emerging topics. In the case of hybrid approaches, Chen (2006) has established research topics using a combination of co-cited relations and term-based techniques in order to analyze emerging trends in the topics of mass extinction and terrorism as well as regenerative medicine (Chen et al., 2012). Similarly, Glänzel, and Thijs (2012) and Reiss et al. (2013) have assembled publications using a combination of bibliographic coupling and term-based approaches.

In addition to these clustering-based approaches, some new methods have also been applied. For instance, Yan (2014) has created research topics using a topic modeling technique in which topics are not created by the aggregation of publications; rather, each publication has a probability distribution over the generated topics.

*Bibliometric Indicators for Identifying Emerging Research Topics*

The next step for identifying emerging research topics is to create indicators according to the definition and attributes of emergence. Our study divides the proposed indicators into five types on the basis of employed analytical methods. The first type is based on the annual number of publications used to detect an emergent trend. For instance, Bengisu (2003) and Small et al. (2014) use growth in the number of publications as an indicator of fast growth; the former further controls for stability after a rapid increase, and the latter also uses the number of publications in the previous time period of an emergent trend as an indicator of radical novelty. The second type of indicators analyze the "core document" in each cluster and tracks it in order to monitor the identification of emerging topics (e.g., Glänzel & Thijs, 2012; Reiss et al., 2013; Small, 2006). The third type explores citation relations within and outside of clusters, analyzes and tracks the position of "leading papers" within clusters to identify emerging topics (e.g., Lee, 2008; Kajikawa & Takeda, 2008; Kajikawa et al., 2008). Other studies have measured the frequency of new terms or keywords as a signal of emerging topics (e.g., Schiebel et al., 2010; Guo et al., 2011; Ohniwa et al., 2010). The last type uses the transition in the number of authors within a research field as an indicator for the detection of emerging topics (e.g., Bettencourt et al., 2008; Guo et al., 2011).

It should be mentioned that most studies combine the previous mentioned indicators in order to explore emerging research topics. For instance, Guo et al. (2011) proposes using the frequency of keywords, the number of authors, and the interdisciplinarity of references as a whole to identify emerging topics.

*Further Discussion on Related Work*

Based on this review of previous literature, limitations on the identification of emerging topics can be summarized as follows. First, in terms of definitions, there is no consensus on the concept of emergence. Various concepts have been used, leading to a variety of indicators being developed. For instance, Cozzens et al. (2010) summarize four major characteristics of emergence: fast recent growth, change to something new, market or economic potential, and increasing "science-basedness", whereas Small et al. (2014) claim that "there is nearly universal agreement on two properties associated with emergence, novelty (or newness) and growth" (p. 1451). Furthermore, as mentioned above, Rotolo et al. (2015) criticize the bibliometric studies on the identification of emergence on the grounds that such approaches tend to focus on measuring the attributes of novelty and fast growth, and to ignore the consideration of other potentially important attributes, such as impact.

Second, in terms of the scope of research, most studies have conducted their analyses using a small database. Some use publications selected from only one or a few specific research field(s) (e.g., Chen, 2006; Kajikawa & Takeda, 2008; Yan, 2014). More specifically, publications under investigation have been retrieved from predefined journals (e.g., Liu et al., 2013) or keywords (e.g., Kajikawa & Takeda, 2008). It is acknowledged that these studies are indeed useful for those studying related fields. Yet, as Small et al. (2014) mention, these studies "cannot identify the currently emerging topics that are of interest to funding bodies and practitioners worldwide" (p. 1450).

Only a few current studies have yet been performed using a large-scale database, mainly those by two research groups, Small and colleagues (Small, 2006; Small et al., 2014; Upham & Small, 2006) and Glänzel and colleagues (Glänzel & Thijs, 2012; Reiss et al., 2013). Two recent large-scale analyses of the identification of emerging topics merit further discussion. One is the ERACEP report (Reiss et al., 2013), the fundamental method of which is described by Glänzel and Thijs (2012); the other is an analysis conducted by Small et al. (2014). The ERACEP report includes WoS publications from 1998 to 2008. The researchers first identify several research fields with a sharp growth rate, and then cluster individual publications of such fields into micro research topics within each five-year time period separately. After matching the clusters of the two different time slices, three paradigmatic cases are said to indicate emerging topics, namely an "existing cluster with an exceptional growth, completely new cluster with its root in other clusters and existing cluster with a topic shift" (Glänzel & Thijs, 2012, p. 404). Another recent study by Small et al. (2014) proposes an approach that combines a direct citation clustering with a co-citation clustering method in a Scopus dataset covering publications over a 15-year period. In this work, a direct citation clustering method is used as the primary source of detection, and a co-citation model is used as a filtering step over the direct citation results. Two properties, novelty and growth, are used to identify emerging topics.

It should be pointed out that the two approaches just discussed employ the technique of matching clusters. To be specific, the approach proposed by Glänzel and Thijs (2012) needs to match the research topics obtained from different time slices by using the core dements in one period and all publications in the other period, while the strategy of Small et al. requires a match of clusters from a direct citation clustering method and a co-citation clustering method. It can be argued that the use of a matching strategy could yield more precise results and could guarantee the coherence of clusters. For instance, Small et al. (2014) mention "the combination of direct citation and co-citation methods used [here] has contributed to this accuracy" (p. 1463). On the other hand, the matching step makes the approaches for identifying emerging topics more complex and less transparent. For instance, understanding in full detail the matching procedures adopted in the two above-mentioned studies is quite challenging. In our opinion, since the effectiveness of any approach for identification is extremely difficult to verify, it is very important that the proposed approaches be transparent, straightforward, and in accord with the perception of emergence. A further discussion of validation can be found at the beginning of the section "Evaluation Tests".

In sum, this study, in an attempt to transcend the limitations mentioned above, adapts the concept defined by Rotolo et al. (2015), develops a more precise definition of an emerging research topic, and puts forward a series of indicators for identifying emerging topics. The effort is also made to establish the linkage between the concept and the constructed method. The methodology and database are introduced in the next section.

**Methodology**

*Definition of Emerging Research Topics*

Rotolo et al. (2015) indicate that the definition and fundamental attributes of an emerging technology are still ambiguous and therefore offer a new definition of an emerging technology:

> [A] radically novel and relatively fast growing technology characterized by a certain degree of coherence persisting over time and with the potential to exert a considerable impact on the socio-economic domain(s) which is observed in terms of the composition of actors, institutions and patterns of interactions among those, along with the associated knowledge production processes. Its most prominent impact, however, lies in the future and so in the emergence phase is still somewhat uncertain and ambiguous. (Rotolo et al., 2015, p.1828)

They further summarize five attributes of an emerging technology: radical novelty, relatively fast growth, coherence, prominent impact, and uncertainty and ambiguity.

By contrast, the present study aims to detect emerging research topics. While there is some overlap between technologies and research topics, they remain distinct concepts. To put it simply, some advances in technologies are not necessarily connected with scientific breakthroughs (Rotolo et al., 2015), whose definition is therefore not entirely appropriate here. For this reason, we adjust the definition of an emerging technology that Rotolo et al. (2015) have proposed in order to accommodate the analysis.

First, the attribute of prominent impact needs to be rethought in order to be relevant to the detection of emerging research topics. Rotolo et al. (2015) use this attribute specifically to refer to the societal and economic impacts of technologies. However, taking into account the differences between science-based research topics and technologies into account, for the former topics, scientific impact should be highlighted rather than societal and economic impact.

Second, uncertainty and ambiguousness is considered another attribute of emerging technologies because Rotolo et al. hold that the societal and economic impact of an emerging technology "lies in the future and so in the emergence phase is still somewhat uncertain and ambiguous" (p.1828). While societal and economic impact is replaced by scientific impact for identifying emerging research topics, it is natural to assume that scientific impact lies in the future as well. However, it should be stressed that scientific impact is more likely to show in a short time period after the emergence of a research topic because researchers tend to present their work at conferences or workshops before it has been officially published and indexed into bibliographic databases. Meanwhile, researchers might easily track the latest developments in their field. Given the above reasons, the scientific impact of a research topic should generally be identified within a relatively short time period after its emergence. Nevertheless, one may argue that the scientific impact of some cases might be unrealized and, then, after a long time period, be noticed and discussed. We acknowledge the existence of such a pattern. However, we claim that if a research topic presents this pattern, it can hardly be considered emerging, since such a pattern does not satisfy our definition on an emerging research topic. Therefore, it is reasonable to require emerging research topics to exert a prominent scientific impact within a rather short time period after its emergence. Hence, the attribute of uncertainty and ambiguousness is of little relevance to the present study.

With these considerations in mind, the following definition of an emerging research topic is proposed here:

> A radically novel and relatively fast growing research topic characterized by a certain degree of coherence, and a considerable scientific impact.

Accordingly, the four attributes of an emerging research topic become radical novelty, relatively fast growth, coherence, and scientific impact. With this definition thus established, the groundwork has been laid to introduce the database and methodology used for the identification of emerging research topics, beginning with their construction.

*Assigning Publications to Research Topics*

Research topics are established based on the clustering method of Waltman and Van Eck (2012, 2013). The reasons for choosing a direct, citation-based clustering method are two-fold. On the one hand, a recent study by Small et al. (2014) has proposed an approach that combines direct citation and co-citation clustering methods. The present study intends to determine whether results similar to those of Small et al. (2014) can be obtained using a less complex methodology that is based on only a direct citation method. On the other hand, Waltman and Van Eck (2012) elaborate the advantages of using direct citation relations in the measurement of the relatedness of publication: "co-citations and bibliographic coupling are more indirect mechanisms than direct citations, and direct citations may therefore be expected to provide a stronger indication of the relatedness of publications" (p. 2380). The use of direct citation relations is also supported by Klavans and Boyack (2015), who conclude that the use of direct citation relations yields more accurate results than bibliographic coupling or co-citation relations.

This study is based on data from the in-house WoS database of the Centre for Science and Technology Studies (CWTS) at Leiden University for the period from 2003 to 2012. All publications of the document type of articles and reviews are included, for a total of around nine million. Figure 1 shows the distribution of the number of publications during the 10 years covered.

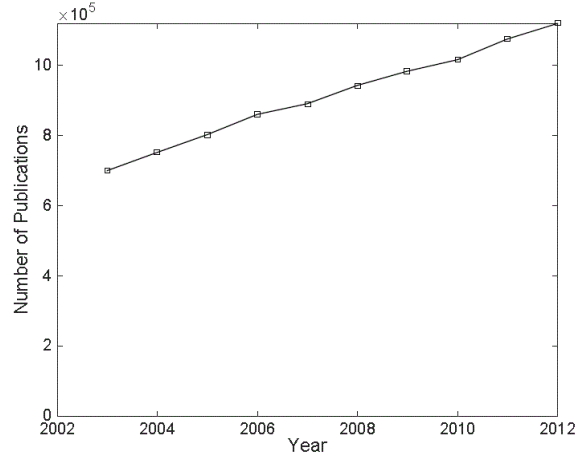

Fig. 1. Number of publications from 2003 to 2012

*Identifying Emerging Research Topics*

In what follows, criteria for identifying emerging topics are proposed and then applied to a specific research topic.

*Defining criteria for identifying emerging research topics.* A research topic can be considered as emerging if it displays the four attributes of radical novelty, relatively fast growth, coherence, and scientific impact. The connections between these attributes and each criterion are elaborated in this subsection, beginning with the attribute of fast growth.

Criterion I – growth. Let $p_{i,t}$ denote the number of publications for research topic $i$ in year $t$, and $p_{i,t,\Delta t}$ the number of publications in the year $t + \Delta t$ in which $\Delta t$ is a time interval. The growth ratio of a topic $i$ in the interval $\Delta t$ is then defined as

$$r_{i,t,\Delta t} = \frac{p_{i,t,\Delta t}}{p_{i,t}}.$$

For a research topic to be considered as emerging, it should show a rapid increase in yearly publications, that is, $r_{i,t,\Delta t} \geq r_{min}$.

It should be mentioned that the annual number of publications is easily influenced by random fluctuations such as the expansion or reduction of the database, which would cause a sudden increase or decrease of yearly publications in a certain research topic. In order to correct for these random fluctuations, we use the smoothed annual number of publications $\bar{p}_{i,t}$ instead of the actual number of publications $p_{i,t}$. This is obtained by calculating the average number of publications in three consecutive years, which can be expressed as

$$\bar{p}_{i,t} = (p_{i,t-2} + p_{i,t-1} + p_{i,t})/3.$$

The growth ratio for a research topic $i$ in the interval $\Delta t$ is actually measured using

$$r_{i,t,\Delta t} = \frac{\bar{p}_{i,t,\Delta t}}{\bar{p}_{i,t}}.$$

Criterion II – novelty. An emerging topic should be novel at the early stage of its emergence, which means that the number of publications in the beginning of a term should be relatively small. Then, for a research topic to be considered as emerging, it should show radical novelty at the early stage of its emergence, that is, $\bar{p}_{i,t} \leq p_{max}$.

Criterion III – scientific impact. An emerging topic should present a prominent scientific impact for which citations are used as a measure. Let $c_{i,t,\Delta t}$ denote the number of citations that publications published between $t$ and $t + \Delta t$ in a research topic $i$ had received in the same time window. This reflects the scientific impact of research topics during a pre-determined time interval. For a research topic to be considered as emerging, it should show a prominent scientific impact, that is, $c_{i,t,\Delta t} \geq c_{min}$.

Criterion IV – coherence. An emerging topic should be coherent. In this study, clusters are generated based on direct citation relations among publications.[1] Furthermore, if the total number of citations received from the publications within a given cluster is less than its total number of publications, then the publications of this cluster may be loosely connected. This can be seen as a sign of the coherence of research topics. Let $h_i$ denote the coherence of a topic $i$, which is measured using the total number of within-cluster citations divided by the total number of publications. For a research topic to be considered as emerging, it should be coherent, that is, $h_i \geq h_{min}$.

In summary, an emerging research topic should satisfy each the following criteria:

1. $r_{i,t,\Delta t} \geq r_{min}$ the research topic should have a rapid growth;

2. $\bar{p}_{i,t} \leq p_{max}$ the research topic should show radical novelty at the early stage of its emergence;

3. $c_{i,t,\Delta t} \geq c_{min}$ the research topic should present a prominent scientific impact; and

4. $h_i \geq h_{min}$ the research topic should be coherent.

It is necessary to give a brief explanation on the role of year $t$ in our criteria. It can be seen that novelty, scientific impact, and scientific impact all relate with $t$. An emerging topic is required to satisfy each of the three criteria for the same value $t$. Of course, it is possible for a research topic to have multiple emergent phases if it satisfies all criteria for multiple values of $t$. However, we record its first emergent period for the convenience of reporting results.

*Applying the criteria for identifying emerging topics.* In this subsection, the cluster regarding graphene research will serve as an example to clarify how these criteria are applied. Its annual number and smoothed number of publications are provided in Table 1.

The parameter value of a time interval needs to be defined at the outset. For instance, if emergence is expected in a relative short term, the parameter value can be set as $\Delta t = 2$, which yields the growth rate at a two-year interval, also shown in Table 1. In the meantime, the smoothed number of publications in the first year of each growth trend can be observed. Next, the attribute of scientific impact is measured calculating its number of citations. For instance, the scientific impact at the year 2007 is measured by summing up the number of citations that the publications published between 2005 and 2007 had received in the same time window. In this way, we obtained the number of citations that the publications had received from 2005 to 2007, which is 6513 in total. Finally, coherence is measured using the total number of within-cluster citations divided by the total publications. In this case, the total number of within-cluster citations is 231,995, which means that the coherence of this research topic is around 21.

Table 1. Statistics on the research topic of graphene ($\Delta t = 2$)

|  | 2003 | 2004 | 2005 | 2006 | 2007 | 2008 | 2009 | 2010 | 2011 | 2012 |
|---|---|---|---|---|---|---|---|---|---|---|
| $p_{i,t}$ | 32 | 38 | 46 | 147 | 415 | 790 | 1197 | 1963 | 2693 | 3422 |
| $\bar{p}_{i,t}$ | - | - | 39 | 77 | 203 | 451 | 801 | 1317 | 1951 | 2693 |
| $r_{i,t,\Delta t}$ | - | - | - | - | 5.2 | 5.9 | 4.0 | 2.9 | 2.4 | 2.1 |
| $c_{i,t,\Delta t}$ |  |  |  |  | 6513 | 16,560 | 29,743 | 48,716 | 70,623 | 86,470 |

All of the information necessary to identify an emerging research topic is thus obtained. Suppose, for example, that the parameter values are set as $r_{min} = 5$, $p_{max} = 100$, $c_{min} = 2{,}500$, and $h_{min} = 1$, this research topic regarding graphene can be considered as emerging. By contrast, when the parameter value of growth is set at $r_{min} = 10$, it cannot be considered as emerging. Thus, whether a topic can be considered as emergent depends on the choice of the parameter values.

**Results**

*Clustering Results*

In this study, around 10,000 clusters are obtained. Each cluster contains 956 publications on average. Figure 2 shows the distribution of publications over clusters. As can be seen, some clusters an extremely small number of publications.[2] Small clusters have little influence on the following analysis and were therefore not excluded.

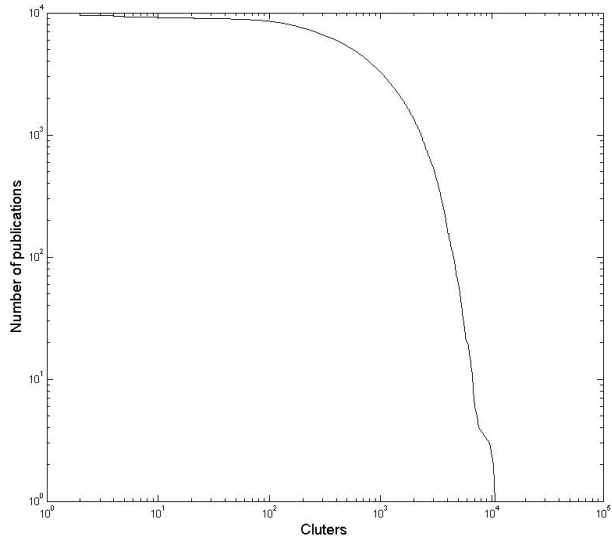

FIG. 2. Distribution of publications over clusters in log-log scale

*Identified Emerging Research Topics*

This analysis uses two sets of parameter values for examining the proposed approach. The use of multiple parameter values can provide insight into the sensitivity of the results. To present a distinction, the parameter values of the time interval are $\Delta t = 2$ and $\Delta t = 5$, respectively. Statistics regarding the attributes of emergence are obtained in the following way. We first determine the maximum growth rate for each topic. Based on it, the smoothed number of publications in the first year of the maximum growth trend can be tracked, and the number of citations that the publications published from $t$ to $t + \Delta t$ had received in the same time window can be calculated. The attribute of coherence is not relevant to a time interval, and it can be easily measured using the total number of within-cluster citations divided by the total number of publications. In this way, we obtained the value of growth, novelty, scientific impact, and coherence for each cluster. The statistics on these attributes are shown in Table 2.

Table 2. Statistics on the attribute of emergence

| Set | Time interval | Attribute | Avg. | Max. | Min. | Std. |
|---|---|---|---|---|---|---|
|   |   | Coherence | 4.57 | 27.78 | 0.5 | 2.79 |
| 1 | $\Delta t = 2$ | Growth | 1.34 | 21.50 | 0.60 | 0.61 |
|   |   | Novelty | 82.28 | 670 | 0.33 | 85.60 |
|   |   | Scientific impact | 972.29 | 34,978 | 0 | 1721.46 |
| 2 | $\Delta t = 5$ | Growth | 1.41 | 57.51 | 0.04 | 1.04 |
|   |   | Novelty | 79.26 | 633.67 | 0.33 | 80.32 |
|   |   | Scientific impact | 4463 | 150,757 | 0 | 7518.60 |

Balancing the statistics of the attributes of emergence, the two sets of parameter values are used, as is summarized in Table 3. The parameter values are relatively strict, apart from coherence. For the first set, topics are selected that both experienced a growth rate of no lower than two and had no fewer than 1,500 citations during a two-year time period. For the second set of parameter values, topics are selected that both presented a growth rate of no less than five and had no fewer than 2,500 citations. It should be noted that the attributes of novelty and coherence do not relate to a time interval, namely $\Delta t$. Thus, for both sets, topics are selected that had no more than 100 publications in the first year of emergence and presented coherence no less than one. It could be argued that the parameter value of coherence is quite small, but the purpose of this attribute is to guarantee that the cluster is not very loosely connected. The two sets of parameter values identified 16 and 15 research topics, respectively, and 12 of which were identified from both sets.

Table 3. Parameter values and the number of identified research topics

|   | Set - 1 | Set - 2 |
|---|---|---|
| $\Delta t$ | 2 | 5 |
| $r_{min}$ | 2 | 5 |
| $p_{max}$ | 100 | 100 |
| $c_{min}$ | 1500 | 2500 |
| $h_{min}$ | 1 | 1 |
| No. of identified research topics | 16 | 15 |

There is a trade-off between precision and recall with this approach, even though, as Small et al. (2014) indicate, it is impossible to estimate the magnitude of the trade-off since a definitive list of emerging topics is lacking. This study identified around 15 emerging research topics using both sets of parameter values. There may be more emerging topics in the 10-year time period. The parameter values established in this study are relatively rigorous, for which reason only a limited number of research topics is identified.

Furthermore, it is difficult to label clusters because some cover a large number of publications. However, for readability purposes we add a brief label to each cluster. More sufficient description can be found in Table A1, where the 10 most frequent terms and the two most cited publications are provided. The first column of Table A1 shows the set of parameter values according to which research topics are identified.

**Evaluation Tests**

The justification of the effectiveness of the proposed methodology requires considerable discussion. Klavans and Boyack (2015) state the following:

> In most fields of science, accuracy is of paramount concern. Admittedly, some fields lend themselves more to accuracy than others. This is particularly true for those fields where physical properties can be measured, those for which gold standards exist, and those where a great deal of research is replicated. Unfortunately, none of these conditions are extant when it comes to the delineation of topics, or the creation of taxonomies of the scientific literature… (p. 988)

It can be seen that no "gold standards" exist in certain bibliometric studies. Small et al. (2014) similarly note the absence of a definitive list of emerging topics, which implies that no widely recognized criteria are available to validate the methods for detecting emerging topics.

Theoretically, the effectiveness of the proposed approach could be validated by replicating previous methods and presenting a systematic comparison, but this strategy is problematic for two reasons. First, the definition and attributes of an emerging topic used in previous studies vary, as explained in "Related Work". Therefore, on account of the differing definitions and attributes, emerging topics identified in these studies could be inconsistent. Second, there is no single optimal level of aggregation, which suggests that different aggregated levels can be used to conduct the identification of emerging topics. Analyses at different aggregated levels also create obstacles in conducting comparisons with other relevant methods.

Further, expert opinions might be used as an alternative means to justify the accuracy of results, even though this approach is problematic as well. Because the present study identifies emerging topics across the entire science system, it would be challenging for experts to assess so broad a scope. Experts may also have different understandings of the concept of emergence and might not accept the definition used in this paper. As a consequence, the criteria that experts use to justify accuracy could differ from the indicators designed here for identifying emerging topics. In sum, while expert opinions indeed represent a valuable way for laymen to gain insight into an unfamiliar research topic, the flexibility of using this approach to validate our results is arguable.

Given that performing a validity test presents difficulties, it is even more important to ensure that the proposed method be straightforward and transparent and to make it accord intuitively with the perception of an emerging research topic. In other words, the proposed indicators should be in strict accordance with the definition and attributes of what qualifies as emergent. In the following subsections, the research topics that have been identified as satisfying the criteria of emerging topics are presented. Next, the research topics detected in the present work are compared with those mentioned in similar studies. The reason for conducting a brief comparison lies in the fact that while we may not yield sufficient evidence for validating our result by comparing with previous work, it can still provide insight into how our results cohere with previous analyses on the identification of emerging research topics. Finally, an in-depth analysis of the field of library and information science (LIS) is presented. These considerations are offered as proof of the effectiveness of the proposed method.

*Demonstrating the Proposed Method*

In this study, the four attributes of an emerging research topic are radical novelty, relatively fast growth, coherence, and scientific impact. Accordingly, Table 4 presents the attributes of emergence for each of the research topics identified. Furthermore, the emergence, development, and disappearance of a research topic can also be observed in terms of yearly publications. Figure 3 illustrates the curve of annual publications for these identified topics.

Table 4. Attributes of emergence for identified topics

| Set | ID | Label | Total no. of publications | Begin year | End year | Novelty | Growth | Coherence | Scientific impact |
|---|---|---|---|---|---|---|---|---|---|
| 1,2 | 1 | graphene | 10,743 | 2005 | 2007 | 38.67 | 5.24 | 21.59 | 6513 |
|     |   |          |        | 2005 | 2010 | 46    | 34.05 |       | 100,340 |
| 1,2 | 2 | superconductivity | 3628 | 2006 | 2008 | 41.33 | 8.26 | 20.79 | 4221 |
|     |   |          |        | 2005 | 2010 | 15    | 40.74 |       | 38,823 |
| 1,2 | 3 | cancer stem cell | 3239 | 2005 | 2007 | 62.67 | 2.45 | 16.82 | 4932 |
|     |   |          |        | 2005 | 2010 | 93    | 6.98  |       | 42,205 |
| 1,2 | 4 | rna seq | 2663 | 2007 | 2009 | 48 | 3.19 | 10.80 | 6784 |
|     |   |          |        | 2005 | 2010 | 47 | 7.31 |       | 23,354 |
| 1,2 | 5 | endoscopic surgery | 2192 | 2006 | 2008 | 39 | 2.71 | 10.31 | 2029 |
|     |   |          |        | 2005 | 2010 | 37 | 8.64 |       | 11,951 |
| 1,2 | 6 | topological insulator | 1999 | 2007 | 2009 | 43.67 | 2.16 | 15.53 | 1894 |
|     |   |          |        | 2005 | 2010 | 30 | 6.63 |       | 9728 |
| 1,2 | 7 | solar cell | 1047 | 2009 | 2011 | 66.33 | 2.40 | 7.10 | 2535 |
|     |   |          |        | 2006 | 2011 | 41 | 5.02 |       | 6238 |
| 1,2 | 8 | SHELX | 5324 | 2006 | 2008 | 78.33 | 4.12 | 2.43 | 4551 |
|     |   |          |        | 2005 | 2010 | 67.33 | 16.53 |       | 24,182 |
| 1,2 | 9 | biomass | 1759 | 2008 | 2010 | 83.33 | 2.30 | 11.18 | 3262 |
|     |   |          |        | 2006 | 2011 | 56.33 | 5.65 |       | 14,633 |
| 1,2 | 10 | resistance | 2200 | 2006 | 2008 | 62.67 | 2.18 | 10.27 | 1736 |
|     |   |          |        | 2005 | 2010 | 45.33 | 5.82 |       | 10,098 |
| 1,2 | 11 | visfatin | 752 | 2006 | 2008 | 29 | 2.20 | 11.67 | 1750 |
|     |   |          |        | 2005 | 2010 | 18 | 5.72 |       | 7854 |
| 1,2 | 12 | transformation optic | 1224 | 2006 | 2008 | 19.67 | 4.49 | 13,92 | 2750 |
| | | | | 2005 | 2010 | 12.33 | 15.43 | | 11,282 |
| 1 | 13 | non-small-cell lung cancer | 579 | 2010 | 2012 | 49 | 2.13 | 10.79 | 3253 |
| 1 | 14 | water splitting | 738 | 2009 | 2011 | 47.33 | 2.15 | 8.10 | 2324 |
| 1 | 15 | thin film | 378 | 2009 | 2011 | 22 | 2.30 | 9.47 | 2062 |
| 1 | 16 | xmrv | 2831 | 2005 | 2007 | 91.67 | 2.12 | 10.82 | 2932 |
| 2 | 17 | genome sequence | 1755 | 2005 | 2010 | 71 | 5.54 | 8.07 | 17,716 |
| 2 | 18 | type 2 diabetes | 441 | 2005 | 2010 | 7 | 6.76 | 8.65 | 3250 |
| 2 | 19 | cognitive radio | 2153 | 2005 | 2010 | 27 | 10.8 | 4.95 | 4429 |

By way of further elaboration, consider, for example, Topic 1, which relates to the electronic properties of graphene research. According to the second most often-cited paper associated with this research topic, by Geim and Novoselov (2007), "[g]raphene is a rapidly rising star on the horizon of materials science and condensed-matter physics. This strictly two-dimensional material exhibits exceptionally high crystal and electronic quality, and, despite its short history, has already revealed a cornucopia of new physics and potential applications" (p. 183). In our analysis, Topic 1 is detected regardless of the set of parameter values used. To be specific, the growth rate of this topic is 5.24 from 2005 to 2007 and reaches 34.05 when its time interval is expanded to a five-year interval. During the first year (2005) in which this topic begins to increase, it only has 39 smoothed publications. In terms of scientific impact, the number of citations for the publications between 2005 and 2007 is over 6,000, and during a five-year time period, the number increases to over 100,000. Meanwhile, this topic is represented by a quite closely associated cluster, the coherence of which is 21. Cluster 1 displays a steep increase in annual publications from 2003 to 2012, as shown in Figure 3. In sum, Topic 1 is indeed a radically novel, relatively fast-growing, and cohesive topic, and it also creates significant scientific impact. Having satisfied the attributes of emergence, it is reasonable to consider the topic of graphene research as emerging.

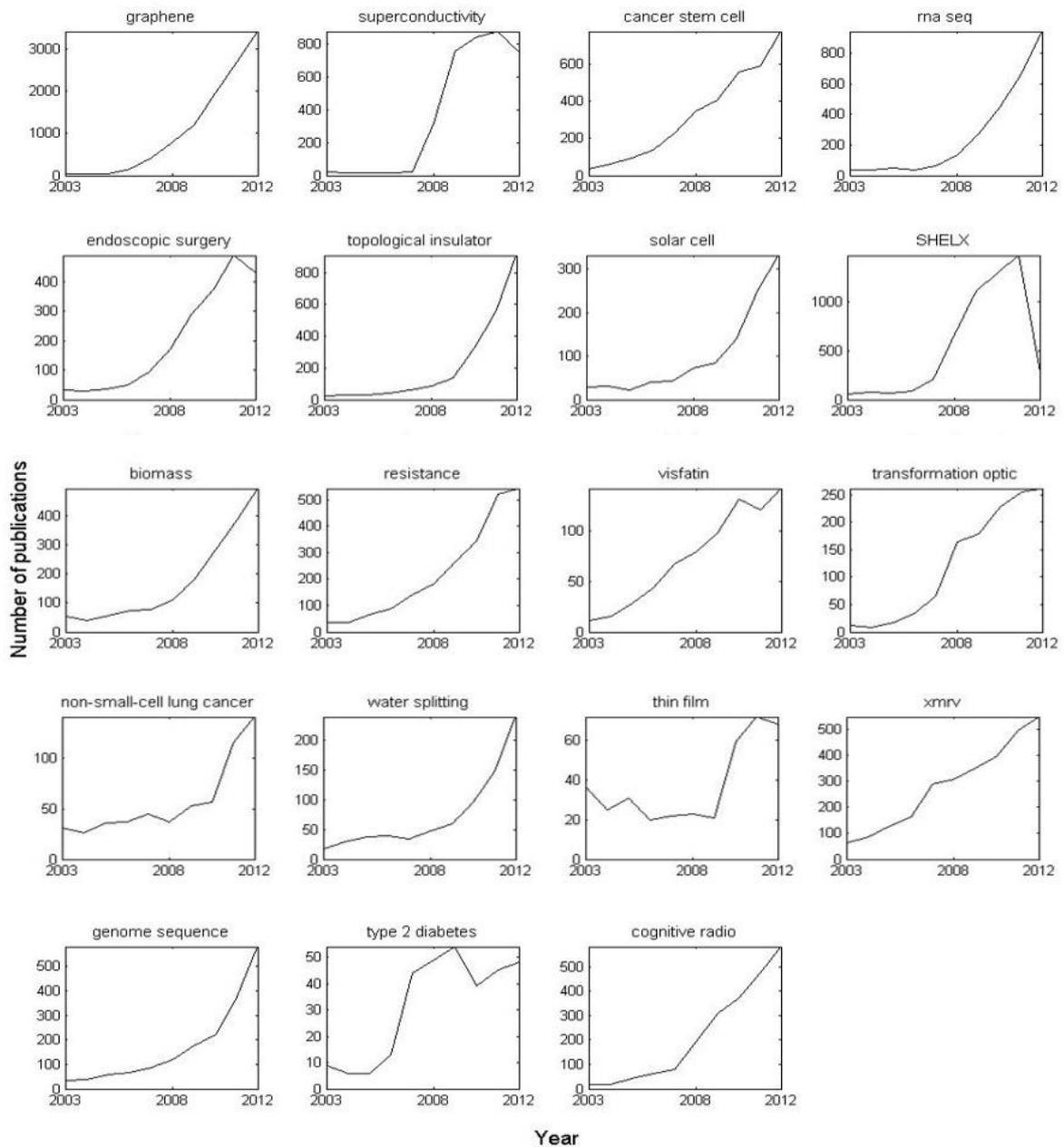

FIG. 3. The trend of publications for the identified topics

It should be mentioned that certain topics are identified using one set of parameter values but cannot be identified using another set, as can be illustrated with the following example. Topic 13, which relates to the studies on EML4-ALK fusion gene in non-small cell lung cancer, is identified as emerging using the first set of parameter values. It displays a sharp increase in yearly publications from 2010 to 2012, as shown in Figure 3. Furthermore, Topic 13 also fits the criteria of novelty, scientific impact and coherence, as shown in Table 4. However, it is not identified as emerging when the second set of parameter values is used, since its highest growth rate in a five-year time interval is only 2.28, which does not satisfy the parameter, namely, $r_{min} = 5$. This is why Topic 13 is not detected as an emerging research topic using a five-year time interval.

The information for the remaining topics that have been identified can be found in Table 4 and Figure 3. These topics will not be demonstrated individually here, but the results indicate that it is reasonable to consider them as emerging.

*Comparison with the Previous Studies*

As discussed in the beginning of this section, comparing with previous analyses on the detection of emerging topics may not provide sufficient evidence for validating our results. In spite of this, a brief comparison can still provide useful information. Therefore, in this subsection, the identified topics are further verified by a comparison with the results from previous scientific studies and other types of documents, such as reports. In addition, it should again be stressed that the aim here is not to systematically examine and compare the effectiveness and accuracy of various methodologies, but rather to offer insight into how the present results cohere with previous analyses.

Previous work on the identification on emerging topics were searched for on Google by combining keywords of our identified topics and terms associated with emergence, which are '*emerging*', '*hot*', '*front*', and '*fast growth*'. Topic 1 serves as an example to illustrate this point further. We use several keywords of Topic 1, such as '*graphene oxide*', '*graphene nanoribbon*', and '*bilayer graphene*' as well as terms related with emergence, such as those just mentioned. However, this process actually leads to a sizable number of publications. The reason could be that researchers tend to use exaggerated positive terms to describe the outcomes and impact of their work (Vinkers et al., 2015). It proves that only the publications aimed at identifying and analyzing emerging topics can provide useful evidence. Based on this type of literature, it finally appears that many studies on analyzing emerging topics have used graphene research as a case for elaborating the attributes of emergence (e.g., Boyack et al., 2014; Klincewicz, 2016). More specific, Small et al. (2014) identify eight research topics with the term '*graphene'* in the labels as emerging. It is possible that Topic 1 has been separated into several clusters in the study of Small et al. (2014), since their analysis was conducted at the micro-level of aggregation, namely, around 85,000 clusters.

Regarding the rest of the research topics identified in the present study, Table 5 shows those also identified in previous works. However, only a few of these works identify emerging research topics from a global perspective. The two main references are the paper by Small et al. (2014) and the information provided by ScienceWatch, which belongs to Thomson Reuters. ScienceWatch conducts analyses based on the publications selected by Essential Science Indicators, and provides various types of analyses of results, for instance, emerging research fronts, fast-moving fronts, and top topics. Unfortunately, the definition and methodology behind these indicators are insufficiently clear, making it difficult to distinguish these entangled terms. It is nevertheless assumed that the different terms mentioned in ScienceWatch can be considered as emerging.

Table 5. Comparison with previous studies on emerging topics

| ID | Label | Identified by other studies |
|---|---|---|
| 1 | graphene | Adams J. & Pendlebury D. (2011). Global Research Report: Materials science and technology. Online: http://strategiprocessen.stratresearch.se/Documents/Strategiprocessen/grr-materialscience.pdf <br> Boyack, K. W. et al. (2014). Characterizing the emergence of two nanotechnology topics using a contemporaneous global micro-model of science. *Journal of Engineering and Technology Management* <br> Small, H. et al. (2014). Identifying emerging topics in science and technology. *Research Policy* <br> Klincewicz, K. (2015). The emergent dynamics of a technological research topic: the case of graphene. *Scientometrics* |
| 2 | superconductivity | Small, H. et al. (2014). Ibid <br> ScienceWatch. (2009). What's new in research <br> Online: http://webcache.googleusercontent.com/search?q=cache:YukLtxj-ahMJ:archive.sciencewatch.com/sciencewatch/dr/sci/09/nov22-09_3+&cd=6&hl=en&ct=clnk&gl=dk |
| 3 | cancer stem cell | Small, H. et al. (2014). Ibid <br> ScienceWatch. (2011). What's hot in medicine <br> Online: http://webcache.googleusercontent.com/search?q=cache:YlHUd37yyFcJ:archive.sciencewatch.com /ana/hot/med2011/+&cd=1&hl=en&ct=clnk&gl=dk |
| 4 | rna seq | Small, H. et al. (2014). Ibid <br> ScienceWatch. (2009). DNA and RNA Sequencing <br> Online: http://webcache.googleusercontent.com/search?q=cache:exlxa3OZlHsJ:sciencewatch.com/dr/tt/2009/09-aprtt-MOL/+&cd=1&hl=en&ct=clnk&gl=dk |
| 5 | endoscopic surgery | Stafinski, T. et al. (2010). The role of surgeons in identifying emerging technologies for health technology assessment. *Canadian Journal of Surgery* |
| 6 | topological insulator | ScienceWatch. (2011). What's hot in Physics <br> Online: http://webcache.googleusercontent.com/search?q=cache:mKULR_YCes8J:archive.sciencewatch.com /ana/hot/phy2011/+&cd=1&hl=en&ct=clnk&gl=dk |
| 7 | solar cell | ScienceWatch. (2008). Emerging research fronts: thin-film organic solar cells <br> Online: http://webcache.googleusercontent.com/search?q=cache:1Tud2CmLtN4J:sciencewatch.com /dr/erf/maps/08apr_phy/+&cd=2&hl=en&ct=clnk&gl=dk |
| 8 | SHELX | Small, H. et al. (2014). Ibid |
| 9 | biomass | Glänzel, W., & Thijs, B. (2012). Using 'core documents' for detecting and labelling new emerging topics. *Scientometrics* |
| 10 | resistance | Chen, A. et al. (Eds.). (2014). Emerging Nanoelectronic Devices. John Wiley & Sons. |
| 11 | visfatin | Choi, S. H. et al. (2013). Clinical implications of adipocytokines and newly emerging metabolic factors with relation to insulin resistance and cardiovascular health. *Frontiers in endocrinology* |
| 12 | transformation optic | Small, H. et al. (2014). Ibid |

| 13 | non-small-cell lung cancer | Sharp D.W. (2012). Medicine top ten: New treatment for at-risk patients with aortic stenosis. |
| | | Online: http://sciencewatch.com/articles/medicine-top-ten-new-treatment-risk-patients-aortic-stenosis |
| 14 | water splitting | ScienceWatch. (2011). Research front map: Photocatalytic hydrogen production. |
| | | Online: http://webcache.googleusercontent.com/search?q=cache:k1n4pMzC3qAJ:archive.sciencewatch.com /dr/rfm/+&cd=2&hl=en&ct=clnk&gl=dk |
| 15 | thin film | Scientists achieve major breakthrough in thin-film magnetism |
| | | Online: http://phys.org/news/2015-08-scientists-major-breakthrough-thin-film-magnetism.html |
| | | Unusual discovery in thin film magnetism |
| | | Online: https://www.sciencedaily.com/releases/2015/08/150813150504.htm |
| | | Chen, A. et al. (Eds.). (2014). Ibid |
| 16 | xmrv | ScienceWatch. (2012) Fast breaking papers |
| | | Online: http://webcache.googleusercontent.com/search?q=cache:MetMLSeUPdsJ:archive.sciencewatch.com /dr/fbp/+&cd=2&hl=en&ct=clnk&gl=dk |
| 17 | genome sequence | Emerging topics in physical virology. Vol. 2. London: Imperial College Press, 2010 |
| 18 | type 2 diabetes | Choi, S. H. et al. (2013). Ibid |
| 19 | cognitive radio | Small, H. et al. (2014). Ibid |
| | | Sciencewatch. (2010). Top Topics |
| | | Online: http://webcache.googleusercontent.com/search?q=cache:PSLBZtWI5sYJ:archive.sciencewatch.com /dr/tt/2010/10-juntt/+&cd=1&hl=en&ct=clnk&gl=dk |

*In-depth Analysis for the Field of LIS*

In this section, research topics relevant to studies in LIS serve as an example for conducting an in-depth analysis since readers of this study are likely to be familiar with this field. This analysis can also contribute to further examine whether the proposed criteria for identifying emerging topics are appropriate and generate meaningful results.

The first step is to extract the LIS-related clusters from our established research topics, and the WoS subject categories are helpful in this respect. A cluster can be regarded as a LIS-related research topic, meaning that more than 50 percent of publications belong to the subject category of LIS. In this way, nine clusters are selected from among all research topics. The frequent terms and the two most-cited publications of these clusters are provided in Table A2. As with the time intervals that are set in previous sections, two intervals are used here, $\Delta t = 2$ and $\Delta t = 5$. Table 6 shows the attributes of emergence for these topics.

As can be seen from Table 6, when we set $\Delta t = 2$, three research topics have growth rates that are slightly greater than the mean value of maximum growth rates, namely, 1.34, which are Topics 90, 7266 and 9354. Among them, Topic 9354 has the highest growth rate, but it is a loosely connected cluster. The growth rates of the other two research topics are lower than our parameter value of growth, that is, $r_{min} = 2$. This implies that no LIS-related topics have a relatively fast growth when using a two-year interval. Regarding scientific impact, only Topic 90 presents a relatively strong impact. However, Topic 90 can hardly be regarded as a radically novel topic, since its number of publications at the beginning year of its emergence is higher than 200.

When $\Delta t = 5$, the three above-mentioned research topics still show fast growth rates higher than the mean value. In fact, the growth rate of Topic 90 is even greater than the 90th growth percentile, that is, $r = 2$. In addition, this topic shows a strong scientific impact from 2005 to 2010. It is not considered as emerging in the previous analysis because of its lack of novelty that the number of publications in 2005 is 327, which exceeds the parameter value for novelty.

Table 6. Statistics on the LIS related research topics

| $\Delta t$ | ID | Total no. of publications | Begin year | End year | Novelty | Growth | Coherence | Scientific impact |
|---|---|---|---|---|---|---|---|---|
| 2 | 90 | 4626 | 2005 | 2007 | 263 | 1.38 | 5.46 | 2139 |
|  | 692 | 2670 | 2006 | 2008 | 262 | 1.11 | 2.73 | 869 |
|  | 2236 | 1411 | 2005 | 2007 | 129 | 1.19 | 1.96 | 436 |
|  | 3208 | 1013 | 2008 | 2010 | 90 | 1.14 | 3.39 | 332 |
|  | 4564 | 650 | 2005 | 2007 | 56 | 1.25 | 1.74 | 122 |
|  | 4974 | 567 | 2007 | 2009 | 56.33 | 1.09 | 4.02 | 296 |
|  | 7266 | 223 | 2005 | 2007 | 14 | 1.45 | 2.37 | 83 |
|  | 7352 | 214 | 2005 | 2007 | 23.33 | 1.17 | 1.18 | 27 |
|  | 9354 | 5 | 2006 | 2008 | 0.33 | 3 | 0.8 | 0 |
| 5 | 90 | 4626 | 2005 | 2010 | 263 | 2.12 | 5.46 | 15,462 |
|  | 692 | 2670 | 2005 | 2010 | 253 | 1.09 | 2.73 | 4363 |
|  | 2236 | 1411 | 2005 | 2010 | 129 | 1.12 | 1.96 | 1713 |
|  | 3208 | 1013 | 2007 | 2012 | 96 | 1.03 | 3.39 | 2233 |
|  | 4564 | 650 | 2005 | 2010 | 56 | 1.29 | 1.74 | 833 |
|  | 4974 | 567 | 2005 | 2010 | 57.67 | 1.02 | 4.02 | 1301 |
|  | 7266 | 223 | 2005 | 2010 | 14 | 1.88 | 2.37 | 467 |
|  | 7352 | 214 | 2005 | 2010 | 23.33 | 0.9 | 1.18 | 194 |
|  | 9354 | 5 | 2006 | 2011 | 0.33 | 3 | 0.8 | 2 |

As can be seen from Table A2, Topic 90 concerns the use of citations for research evaluation, which is an essential subject in bibliometric studies. This does not seem to be a very novel research topic. However, Hirsch proposed an H-index in 2005 that can be used to assess the performance of individual researchers. As Waltman and Van Eck (2012) indicate, "[t]he introduction of the h-index (or Hirsch index) in 2005 has had an enormous influence on bibliometric and scientometric research" (p. 406). Zhang et al. (2011) also conclude that the paper on the H-index has attracted a great deal of interest across the entire scientific community, including the natural and social sciences. It implies that Topic 90 can be regarded as the most emerging topic in the LIS field during the 10-year time period, even though it does not satisfy the two sets of criteria used in this study. This outcome also suggests that the parameter values should be defined based on the specific research purpose and the research field under investigation.

**Discussion and Conclusions**

This study proposed a new methodology for the detection of emerging research topics. The proposed approach was applied to topics that are constructed based on the direct citation relations of individual publications. Using two sets of parameter values, several emerging research topics were identified. The evaluation was performed by demonstrating the proposed approach, and by referring to previous studies and reports regarding the identification of emerging topics. In addition, an in-depth analysis on the field of LIS was conducted to examine this approach further.

The merit of the proposed method lies in the fact that the present study carefully elaborates a definition of an emerging research topic and stresses the linkages between the attributes of emergence and the indicators that have been developed. In addition, the definition and attributes of emerging research topics proposed in this study follow the work of Rotolo et al. (2015) in explaining the relationship between each attribute of emergence and its operational indicator. Thus, the present approach can improve on current bibliometric approaches to identifying emerging topics since it is conceptually straightforward and operationally transparent.

Furthermore, our approach is flexible, which implies that it can be adjusted to satisfy various purposes. For researchers interested in identifying emerging topics in a particular field, the parameter values of the present approach could be re-set based on the field under investigation. Of course, a suitable aggregated level should also be selected to perform the identification of emergence in the target field. More specifically, when researchers are only concerned about emerging research topics in social science studies, certain factors should be taken into account, for instance, the fact that research topics in social science tend to have a relatively small number of publications and are likely to show a slow rate of knowledge renewal. Under such circumstances, it is better to proceed on a more fine-grained aggregation level and to relax the restriction of certain parameter values, such as growth. In short, when researchers are concerned about a particular field, the identification of emerging topics should be performed at a suitable aggregated level with a comprehensive consideration of the characteristics of this field. Moreover, it should be noted that the proposed approach can be applied only once users concur regarding the definition of emergence used in this paper. Potential users must therefore carefully consider whether their understanding of emergence is consistent with the one that is advanced in this study before employing the present approach.

However, some general limitations remain of bibliometric-based studies that identify emerging topics. First, our analysis was based on the publications in the WoS database with the document types of article and review. In this case, the methodology proposed may be less useful for the fields that take other document types, such as monographs and conference papers, as novel knowledge carriers. For instance, few computer science topics were identified as emerging in this paper. The reason could be that frontier research of computer science is mainly presented at conferences. Further, bibliometric approaches are sensitive to fluctuations in the database such as the indexing of new journals associated with a certain research topic, as discussed in "Methodology", which could also influence the accuracy of analyses. Rotolo et al. (2015) suggest that other data sources may be useful for the detection of emerging topics, such as funding information and social media. This may worth studying in future research.

**Footnotes:**

[1] The clustering method used in this paper cannot automatically ensure that the generated clusters are sufficiently coherent. In this case, using a criterion to evaluate the coherence of clusters is necessary.

[2] An alternative approach would have been to exclude small clusters or to merge them with larger clusters. A method for merging smaller clusters with larger ones is described by Waltman and Van Eck (2012).

**Acknowledgements**

This paper was written during a research stay at CWTS, Leiden University. I would like to thank Ludo Waltman for his very helpful discussion and suggestions. I also acknowledge the generous help of Kevin W. Boyack in providing their data. I appreciate the advice from Jos Winnink, Martijn Visser, Ulf Sandström, Ulf Heyman, and Ismael Rafols. In addition, I sincerely thank the editor and two anonymous reviewers for their valuable comments.

# Appendix

Table A1. Identified research topics

| Set | ID | Label | Terms | Publications |
|---|---|---|---|---|
| 1, 2 | 1 | graphene | graphene; graphene oxide; graphene nanoribbon; bilayer graphene; effect; synthesis; epitaxial graphene; graphite; electronic property; electronic structure | Novoselov, K.S. et al. (2004). *Science.* Geim, A. K., & Novoselov, K. S. (2007). *Nature materials.* |
| 1, 2 | 2 | superconductivity | superconductivity; superconductor; iron; single crystal; effect; iron pnictide; fe1 xcox; electronic structure; pressure; magnetism | Kamihara, Y. et al. (2008) *Journal of the American Chemical Society.* Rotter, M. et al. (2008). *Physical Review Letters.* |
| 1, 2 | 3 | cancer stem cell | cell; cancer; cancer stem cell; expression; stem cell; cd133; identification; cancer stem cells; breast cancer; characterization | Al-Hajj, M et al. (2003). *Proceedings of the National Academy of Sciences.* Singh, S. K, et al. (2004). *Nature.* |
| 1, 2 | 4 | rna seq | analysis; next generation sequencing; rna seq; sequencing; assembly; identification; exome sequencing; cancer; application; detection | Mortazavi, A. et al. (2008). *Nature methods.* Li, H. et al. (2009). *Bioinformatics.* |
| 1, 2 | 5 | endoscopic surgery | note; natural orifice transluminal endoscopic surgery; initial experience; laparoendoscopic single site surgery; porcine model; comparison; natural orifice transluminal endoscopic surgery; case; single incision laparoscopic cholecystectomy; single incision laparoscopic surgery | Kalloo, A. N. et al. (2004). *Gastrointestinal endoscopy.* Marescaux, J. et al. (2007). *Archives of surgery.* |
| 1, 2 | 6 | topological insulator | topological insulator; surface; majorana fermion; dimensional topological insulator; spin orbit; optical lattice; bose einstein condensate; graphene; state; effect | Hasan, M. Z., & Kane, C. L. (2010). *Reviews of Modern Physics.* Bernevig, B. A. et al. (2006). *Science.* |
| 1, 2 | 7 | solar cell | solar cell; thin film silicon solar cell; light; thin film solar cell; effect; thin film; enhancement; performance; influence; absorption enhancement | Atwater, H. A., & Polman, A. (2010). *Nature materials.* Pillai, S. et al. (2007).*Journal of applied physics.* |
| 1, 2 | 8 | SHELX | methyl; bis; chlorophenyl; phenyl; dimethyl; ethyl; chloro; crystal structure; methylphenyl; methoxyphenyl | Sheldrick, G. M. (2007). *Acta Crystallographica Section A: Foundations of Crystallography.* Spek, A. L. (2009). *Acta Crystallographica Section D: Biological Crystallography.* |

| | | | |
|---|---|---|---|
| 1, 2 | 9 | biomass | glycerol; cellulose; conversion; synthesis; catalyst; biomass; fructose; production; glucose; hydrogenolysis | Corma, A. et al. (2007). *Chemical Reviews.* Zhao, H. et al. (2007). *Science.* |
| 1, 2 | 10 | resistance | resistive; effect; resistance; characteristic; thin film; behavior; resistive switching; mechanism; memory; property | Waser, R., & Aono, M. (2007). *Nature materials.* Strukov, D. B. et al. (2008). *Nature.* |
| 1, 2 | 11 | visfatin | visfatin; effect; type; patient; chemerin; expression; factor; obesity; insulin resistance; association | Fukuhara, A. et al. (2005). *Science.* Berndt, J. et al. (2005). *Diabetes.* |
| 1, 2 | 12 | transformation optic | transformation optic; metamaterial; design; transformation; cloak; invisibility; transformation medium; invisibility cloak; application; light | Pendry, J. B. et al. (2006). *Science.* Schurig, D. et al. (2006). *Science.* |
| 1 | 13 | non-small-cell lung cancer | anaplastic lymphoma kinase; alk; anaplastic large cell lymphoma; crizotinib; identification; npm alk; non small cell lung cancer; neuroblastoma; small cell lung cancer; large cell lymphoma | Soda, M. et al. (2007). *Nature.* Kwak, E. L. et al. (2010). *New England Journal of Medicine.* |
| 1 | 14 | water splitting | alpha; water; thin film; effect; photoelectrochemical water splitting; water oxidation; characterization; photoelectrochemical property; synthesis; water splitting | Kay, A. et al. (2006). *Journal of the American Chemical Society.* Kanan, M. W., & Nocera, D. G. (2008). *Science.* |
| 1 | 15 | thin film | thin film; effect; magnetic property; ceramic; structure; bifeo3; film; ferroelectric property; synthesis; electrical property | Wang, J. et al. (2003). *Science.* Eerenstein, W. et al. (2006). *Nature.* |
| 1 | 16 | xmrv | xmrv; virus; xenotropic murine leukemia virus; chronic fatigue syndrome; rnase l; prostate cancer; evidence; xenotropic murine leukemia virus related virus; role; infection | Urisman, A. R. J. M. et al. (2006). *PLoS Pathog.* Lombardi, V. C. et al. (2009). *Science.* |
| 2 | 17 | genome sequence | complete genome sequence; genome sequence; draft genome sequence; isolated; analysis; sequence; metagenomic; genome; metagenome; metagenomic analysis | Venter, J. C. et al. (2004). *Science.* Margulies, M. et al. (2005). *Nature.* |
| 2 | 18 | type 2 diabetes | retinol binding protein; serum retinol binding protein; type; insulin resistance; patient; diabetes; level; association; effect; rbp4 | Yang, Q. et al. (2005). *Nature.* Graham, T. E. et al. (2006). *New England Journal of Medicine.* |
| 2 | 19 | cognitive radio | cognitive radio network; cognitive radio; cognitive radio system; spectrum sensing; cognitive radio networks; cognitive radios; spectrum; cooperative spectrum sensing; ofdm; power allocation | Haykin, S. (2005). *IEEE Journal on Selected Areas in Communications.* Akyildiz, I. F. et al. (2006). *Computer networks.* |

Table A2. Labels and the two most cited publications of LIS-related topics

| ID | Terms | Publications |
|---|---|---|
| 90 | science; research; journal; impact; analysis; bibliometric analysis; citation; h index; publication; comparison | Hirsch, J. E. (2005). *Proceedings of the National academy of Sciences of the United States of America.*<br>Egghe, L. (2006). *Scientometrics.* |
| 692 | information; information literacy; library; web; use; academic library; study; case study; information science; impact | Vakkari, P. (2003). *Annual review of information science and technology.*<br>Borlund, P. (2003). *Journal of the American Society for information Science and Technology.* |
| 2236 | library; use; case study; document supply; university; development; academic library; e book; review; survey | Littman, J., & Connaway, L. S. (2004). *Library resources & technical Services.*<br>Langston, M. (2003). *Library Collections, Acquisitions, and Technical Services.* |
| 3208 | E-government; case; role; case study; government; information; citizen; impact; E- government service; lesson | West, D. M. (2004). *Public administration review.*<br>Norris, D. F., & Moon, M. J. (2005). *Public administration review.* |
| 4564 | web; user; library; image; use; metadata; analysis; information; study; taxonomy | Choi, Y., & Rasmussen, E. M. (2003). *Journal of the American society for information science and technology.*<br>Antelman, K. et al. (2006). *Information technology and libraries.* |
| 4974 | web; study; link; comparison; case; information; south Korea; science; case study; impact; research | Wilkinson, D. et al. (2003). *Journal of information science.*<br>Vaughan, L., & Thelwall, M. (2003). *Journal of the American Society for Information Science and Technology.* |
| 7266 | information system; research; information systems research; year; discipline; latent semantic indexing; law; nature; Canada; reflection | Benbasat, I., & Zmud, R. W. (2003). *MIS quarterly.*<br>Chen, W., & Hirschheim, R. (2004). *Information systems journal.* |
| 7352 | use; Nigeria; digital divide; case study; ict; koha; Africa; sub Saharan Africa; development; information | Adomi, E. E. et al. (2003). *The Electronic Library.*<br>Hoe-Lian Goh, D. et al. (2006). *Online Information Review.* |
| 9354 | Paul Otlet; visualization; knowledge; ontology; society; circle; organize research; theory; interface; design | Van den Heuvel, C. (2009). *Knowledge Organization.*<br>Ducheyne, S. (2005). *Knowledge Organization.* |